\documentclass[%
 aip,
 jmp,%
 amsmath,amssymb,
reprint,%
]{revtex4-1}

\usepackage{graphicx}
\usepackage{dcolumn}
\usepackage{bm}
\usepackage{hyphenat}

\begin{document}

\preprint{AIP/123-QED}

\title[Applied Physics Letters(2015)]{Energy-efficient hybrid spintronic-straintronic reconfigurable bit comparator}

\author{Ayan K. Biswas}
\affiliation{%
Department of Electrical and Computer Engineering, Virginia Commonwealth University, Richmond,
Virginia 23284, USA
}

\author{Jayasimha Atulasimha}%
\affiliation{
Department of Mechanical and Nuclear Engineering, Virginia Commonwealth University, Richmond,
Virginia 23284, USA
}%

\author{Supriyo Bandyopadhyay}
\affiliation{%
Department of Electrical and Computer Engineering, Virginia Commonwealth University, Richmond,
Virginia 23284, USA
}
%

\date{\today}

\begin{abstract}
We propose a reconfigurable bit comparator implemented with a nanowire spin valve whose contacts are magnetostrictive
and possess bistable magnetization. Reference and input bits are ``written'' into the magnetization states of the
two contacts with electrically generated strain and the  spin-valve's resistance is lowered if the bits match.
Multiple comparators can be interfaced in parallel with a magneto-tunneling junction
to determine if an N-bit input stream matches an N-bit reference stream bit by bit. The system is robust against thermal
noise at room temperature and a 16-bit
comparator can operate at $\sim$294 MHz while dissipating at most $\sim$19 fJ per cycle.
\end{abstract}
\maketitle

Digital signal processing employing electron spins to store and process bit information
offers certain advantages over traditional charge-based paradigms \cite{book, wolf}.
Here, we propose a hybrid spintronic-straintronic digital reconfigurable N-bit comparator that can operate at room temperature with reasonable speed while dissipating very little energy.
It is also ``non-volatile'', meaning that the input bits, reference bits and
 the result of the bit stream comparison (i.e. whether the input and reference bit streams match or not) can be stored
 permanently in the magnetization
states of magnets.

\begin{figure}[!ht]%
 \centering
  \includegraphics[width=3.6in]{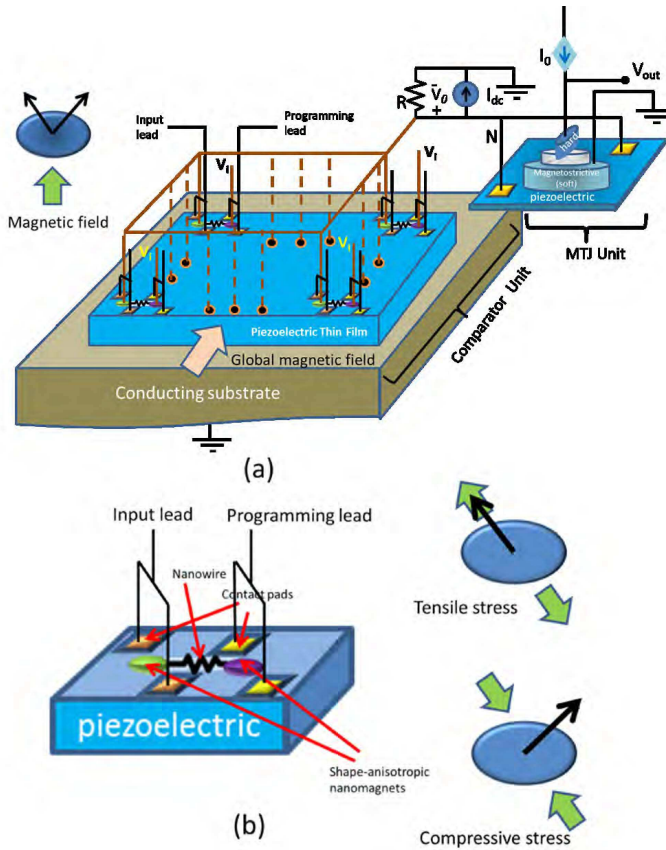}
 \caption[(a) Caption of figure]{(a) A straintronic-spintronic multi-bit comparator integrated with
a magneto-tunneling junction whose resistance indicates whether the input and reference bit streams
match bit by bit. The MTJ unit and the comparator unit share the same (grounded) conducting substrate
although that has not been shown explicitly in the figure for the sake of clarity. (b) [Left panel] A single bit
comparator unit
showing the nanowire
spin valve with magnetostrictive contacts fabricated on a piezoelectric layer. The programming and input leads are shown.
[Right panel] Uniaxial tensile stress applied along one stable orientation of the nanomagnet takes the magnetization
to that orientation while compressive stress takes the magnetization to the other orientation.  }
  \label{fig:comparator}
\end{figure}

Figure 1(a) shows the schematic of an N-bit comparator fabricated on a conducting
n$^{+}$-Si substrate and a piezoelectric layer. A single bit comparator block is shown in the left panel of Fig. 1(b) and consists of a nanowire
``spin valve'' whose two ferromagnetic contacts are essentially two-phase multiferroics made of a magnetostrictive
material elastically coupled to an underlying piezoelectric film. The spacer layer of the spin valve is a semiconductor
nanowire in which electron transport is single-channeled and hence immune to D'yakonov-Perel' spin relaxation
\cite{pramanik}.
Strong suppression of spin relaxation due to single-channeled transport has been
observed in 50-nm diameter electrochemically self assembled InSb nanowires at room temperature \cite{saumil}
and progressive suppression with decreasing width of a nanowire has been reported in InGaAs nanowires \cite{holleitner}. 
To fabricate the spin valve structure, a single $\sim$50 nm diameter 
InSb nanowire can be captured between two lithographically delineated magnetostrictive
contacts (deposited on a piezoelectric film) using dielectrophoresis
\cite{arun}.

The contacts are shaped like elliptical disks and a global static magnetic field
is directed along their minor axes. This makes the magnetization orientation of either contact {\it bistable}.
The two stable orientations lie in two adjacent quadrants in the ellipse's plane between the major and minor axes and
subtend an angle of $\sim$90$^{\circ}$
with each other
\cite{Tiercelin1, Tiercelin2} as shown in the top left corner of Fig. 1(a). Uniaxial tensile stress {\it of sufficient
strength} applied along one of these stable orientations will drive the magnetization of a contact
to that stable orientation while compressive stress of sufficient strength will drive it to the other stable orientation if
the
magnetostriction coefficient of the contact material is positive (the reverse will happen if it is negative)
\cite{Tiercelin1, Tiercelin2}. This is shown in the right panel of Fig. 1(b).

Two pairs of electrically shorted (non-magnetic) electrodes are also delineated on the piezoelectric film, with each pair flanking a
 magnetostrictive contact such that the line joining the centers of the electrodes in a pair is approximately collinear
 with one stable
magnetization orientation of that contact [see left panel of Fig. 1(b)]. Application of
a voltage (of appropriate sign and magnitude) between an electrode pair and the underlying grounded n$^+$-Si substrate
will produce a tensile stress along the line joining the electrode centers and a compressive stress in the in-plane direction
perpendicular
to that line \cite{Lynch}. Voltage of the opposite polarity
will interchange the signs of the stresses
\cite{Lynch}. This stressing scheme requires certain geometrical constraints to be imposed on electrode size and separation,
which are described in ref. [\onlinecite{Lynch}] and not discussed here. In the end, by applying a positive voltage at the electrode pair, we can
make the magnetization of the interposed magnetostrictive contact of the spin valve
orient along one stable direction, while by applying a negative voltage, we can make it orient along the
other stable direction.

The way a single bit comparator works is as follows: First, a reference bit is ``written'' and stored in the comparator by
activating the electrode pair surrounding one magnetostrictive contact of the spin valve.
This will orient that contact's magnetization along one of its two stable states. Let us assume that the
state attained when the electrode voltage is positive encodes bit `1' and the other state
(attained when the electrode voltage is negative) encodes bit `0'. Thus, a positive electrode voltage will write and store the bit 1
while a negative voltage will write and store the bit 0.  This electrode pair will become the
``programming lead'' since it programs the stored (reference) bit. Since the stored bit can be always ``re-programmed''
with the programming lead, the comparator is {\it reconfigurable}. Next, to carry out the bit comparison, the voltage
encoding the input bit (positive voltage = 1; negative voltage = 0) is applied to the electrode
pair surrounding the other contact of the spin valve
(this electrode pair is therefore called the ``input lead''). If the input and the
stored (reference) bits are the same, then the polarities of the voltages applied at the input lead and
programming lead will have been the same, making the magnetizations of the spin valve's two contacts {\it mutually parallel}.  
This will reduce the spin-valve's electrical resistance. On the other hand, if the input and reference
bits are different, then the magnetizations of the two contacts will be in the two different orientations, which are
roughly  perpendicular, and
the spin valve resistance will be higher. Thus, the spin valve's {\it resistance} is the indicator of match/mismatch
between the input  and reference bits.
The ratio of the resistances indicating mismatch and match is 1 + $\eta_1 \eta_2$
\cite{book} (assuming no spin relaxation
in the spacer layer), where the
$\eta$-s are the spin injection and detection efficiencies of the two contacts. At room temperature,
spin injection efficiencies of $\sim$70\% have been demonstrated \cite{salis}; therefore, this ratio can be
 $\sim$ 1.5. Note that it would have been further lowered if there was significant spin relaxation in the spacer, which is
 why it is important to suppress relaxation by ensuring single channeled transport in the spacer layer.

The magneto-tunneling junction (MTJ) unit in Fig. 1(a) is fabricated on the same
(grounded) conducting substrate and the piezoelectric film as the spin valve comparators. Its role is to determine if
an N-bit input stream and a pre-programmed N-bit reference stream stored in the comparators
match exactly {\it bit by bit}
and store the result (``yes'' or ``no'') in its bistable resistance state.
The
soft layer of the MTJ is an elliptical magnetostrictive nanomagnet in elastic contact with
the underlying piezoelectric layer and it has two stable magnetization orientations that are roughly perpendicular to each
other and lie in the plane of the soft layer in two adjacent quadrants between the
major and minor axes, just like the magnetostrictive contacts of the spin valves. The hard layer
of the MTJ is elliptical with very high eccentricity and its two stable states are roughly along the major
axis of the ellipse because of the very high shape anisotropy. The hard nanomagnet is placed such that
its major axis is collinear with one of the stable magnetization orientations of the soft nanomagnet
(resulting in a ``skewed MTJ stack'' where the major axes of the two nanomagnets are at an angle). The
 hard nanomagnet is then magnetized permanently in a direction that is {\it anti-parallel} to the stable
magnetization direction of the soft magnet. Thus, when the soft nanomagnet is
 in one stable state, the magnetizations of the hard and
soft layers are anti-parallel (high MTJ resistance), while when the soft nanomagnet is in the other stable state,
the magnetizations of the two layers are roughly perpendicular (low MTJ resistance).

There are two contact pads flanking the MTJ and the line joining their centers is aligned along the  magnetization of
the hard layer (which also happens to be collinear with one stable magnetization orientation of the soft layer).
At the beginning of any comparison cycle, the MTJ resistance is ``reset''; the soft layer's magnetization is oriented
in the stable direction that is anti-parallel to that of the hard layer's
by applying a voltage of the appropriate polarity between the contact pads and substrate [using connections not shown
in Fig. 1(a) for the sake of clarity] that generates tensile stress in that direction.   The appropriate polarity depends on the direction of piezoelectric poling
and the sign of the soft layer's magnetostriction coefficient. Once the reset operation is over, the MTJ
is left in the high resistance state.

\begin{figure}[!ht]%
 \centering
  \includegraphics[width=3.5in]{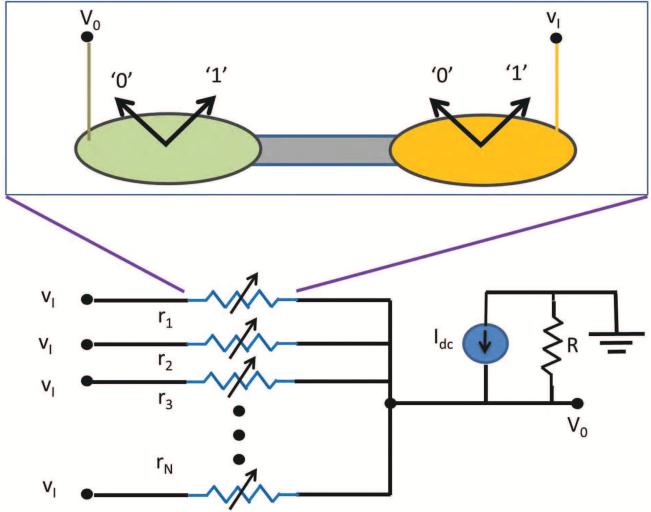}
 \caption[Caption of figure]{Equivalent circuit of the N-bit comparator including the comparison and decision making units.}
  \label{fig:EquivalentCir}
\end{figure}

The way the MTJ unit works to determine match or mismatch between the input and reference bit streams is as follows:
One contact of each spin valve is connected to a common voltage source $V_I$ while the
other contact is connected to one terminal of a passive resistor $R$ whose remaining
terminal is grounded as shown in Fig. 1(a). The corresponding circuit is shown in Fig. 2. Using superposition,
the voltage $V_0$ appearing across the resistor $R$ (and hence across the piezoelectric layer
underneath the MTJ -- see Fig. 1(a)) is found to be
\begin{equation}
V_0 = V_I \sum_{n=1}^{N} {{R_n}\over{R_n + r_n}} ,
\end{equation}
where $R_n = R \parallel r_1 \parallel r_2 \parallel \cdot \cdot \cdot \parallel r_{n-1} \parallel r_{n+1} \parallel \cdot \cdot \cdot \parallel r_{N}$ and $r_n$ is the resistance of the $n$-th
spin valve. The above equation reduces to
\begin{equation}
V_0 \approx V_I \sum_{n=1}^{N} {{R}\over{r_n}} = V_I R G_{\parallel} ~~~ {\tt if} ~ R_n \ll r_n,
\end{equation}
where $G_{\parallel}$ is the sum of the spin valve conductances ($G_{\parallel} = \sum_{n=1}^N {{1}\over{r_n}}$). The strong inequality in the last
equation is realistic since single nanowires tend to have very high resistances.

It is easy to see now that when the input and reference bit streams
match exactly bit by bit, the output voltage $V_0$ is maximum because $G_{\parallel}$ is
maximum. We call this value of $V_0$ ``$V_{match}$''. When
one or more bits do not match, $V_0 < V_{match}$.

Note that the voltage $V_0$ is applied at the MTJ electrodes. There is a threshold positive voltage $V_{th}$ which,
when applied at these electrodes, will generate enough
compressive stress in the soft layer of the MTJ to rotate its magnetization from the initial (``reset'')
orientation to the other stable orientation that is
roughly perpendicular to the magnetization of the hard layer. At this threshold
voltage, the strain energy overcomes the energy barrier between the two stable orientations
to make the switching occur. This will abruptly take the MTJ from the initial (post-reset) high-resistance 
state to the low-resistance state and reduce the resistance by
a factor of $1/\left (1 - \eta_1 \eta_2 \right )$. The MTJ is biased by a constant current source
$I_0$ which generates an output voltage $V_{out} = I_0 R_{MTJ}$, where $R_{MTJ}$ is the MTJ resistance.
If $V_0 \geq V_{th}$, then $V_{out}$ is low (because $R_{MTJ}$ is low); otherwise, $V_{out}$ is high.

We will now set $V_{match} = V_{th}$. This will ensure that $V_{out}$ will be low {\it if and only if}
the bit stream and the reference stream match exactly bit by bit. Otherwise, $V_{out}$ will be high.
Thus, by monitoring $V_{out}$, we can determine if the bit streams match exactly. A comparator of this type has been
proposed earlier by Datta et al. \cite{datta} where the magnets are switched with spin torque and no spin valves are used.

There are two ways to ensure that the equality $V_{match} = V_{th}$ holds.
We can design the soft layer of the MTJ (shape, material and dimensions) as well as the global magnetic field to make
$V_{th}$ satisfy this equality, but it is challenging to control $V_{th}$. An easier way is to fine tune $V_0$ with a
variable
current source $I_{dc}$ as shown in Fig. 1(a). This will add an extra term $I_{dc} R$ to $V_0$
and change the required equality to $V_{match} + I_{dc}R
= V_{th}$. We now do not have to precisely craft $V_{th}$. Instead,
we will tune $I_{dc}$ to enforce the above equality.

All of the above discussion appears relevant only to 0 K temperature when there is no thermal noise to
smear the sharp $V_{th}$. At room temperature, there will be a broadening of the threshold to
$V_{th} \pm \Delta V/2$. Therefore, to make the scheme work at room temperature, we have to ensure that if even
{\it one bit does not match}, the resulting $V_0$ appearing across the resistor $R$ is considerably
{\it less} than $V_{th} - \Delta V/2$. This can be ensured by choosing $V_I$, $R$, $I_{dc}$ and the spin valve resistances
in the low- and high-resistance states judiciously.

\begin{figure}[!ht]%
 \centering
  \includegraphics[width=3.5in]{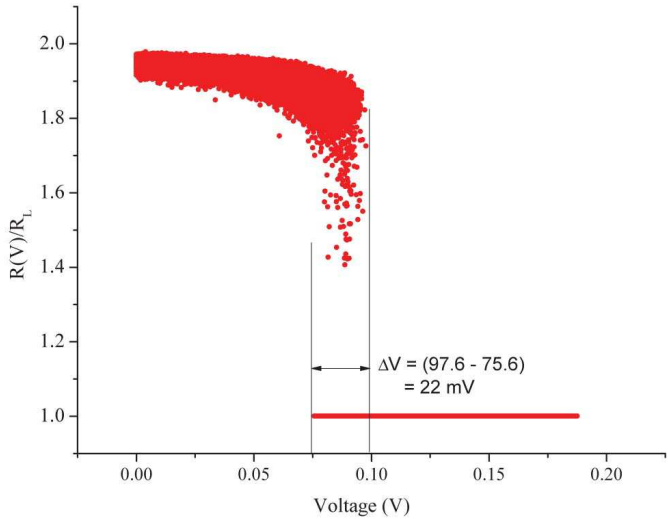}
 \caption[Caption of figure]{Switching characteristic of the MTJ switch $R(V)$ versus $V$
in the presence of thermal noise at room temperature. This plot is generated by simulating 10$^5$ switching trajectories
in the manner of refs.
[\onlinecite{kuntal}] and [\onlinecite{Biswas2014}] to find the thermal spread in the switching threshold.}
  \label{fig:MTJ_LU_300K_voltage}
\end{figure}

First, we need to determine the smearing of $V_{th}$ by simulating the MTJ switching dynamics at room temperature. We assume
that the soft layer of the MTJ is made of Terfenol-D and shaped like an
elliptical disk of major axis 100 nm, minor axis 42 nm and thickness 16.5 nm. These dimensions ensure
that the soft nanomagnet has a single domain \cite{cowburn}  and the energy barrier between its two stable magnetization
states is $\sim$73 kT at room temperature if the global magnetic field is 0.14 Tesla \cite{beleggia}.
The stress generated in the soft nanomagnet
by a given voltage applied at the MTJ's electrode pair is estimated from ref. [\onlinecite{Lynch}]
assuming that the piezoelectric layer is a 100-nm thick lead-zirconate-titanate (PZT) film. We
simulate the magnetodynamics of the soft nanomagnet under stress in the presence of thermal noise in the manner of
refs. [\onlinecite{kuntal}] and [\onlinecite{Biswas2014}], i.e. by simulating switching trajectories according to the
stochastic
Landau-Lifshitz-Gilbert (LLG) equation \cite{spedalieri}. When the soft layer switches from one stable magnetization state
to the other, the MTJ resistance changes by a factor of $1/\left (1 - \eta_1 \eta_2 \right ) \approx 2$
if we assume $\eta_1 = \eta_2 = 0.7$ \cite{salis}. Higher resistance ratios exceeding 6:1 have been demonstrated at room temperature \cite{ikeda},
but we will be conservative and assume the ratio to be 2:1. Fig. 3 shows the resistance $R(V)/R_{low}$ versus
$V$ where $R(V)$ is the MTJ resistance when a voltage $V$ is applied at the electrode pair and $R_{low}$ is the low resistance of the MTJ. To generate this scatter plot, 10$^5$ switching trajectories were
simulated at room temperature. Clearly, room-temperature fluctuations broaden the switching threshold by $\sim$22 mV
(i.e. $\Delta V$ = 22 mV), and switching can occur anywhere between 75.6 and 97.6 mV applied at the MTJ's electrode pair.

We will now have to ensure that when all bits in the input and reference streams match, $V_0 =
V_I R G_{\parallel}^{\tt max} +I_{dc}R
\geq$ 97.6 mV, and when there is just one mismatch, $V_0 = V_I R G_{\parallel}^{\tt 1-mismatch} +I_{dc}R
\leq$ 75.6 mV. We ensure this by choosing $V_I$ = 10 V, $I_{dc}$ = 1.51 $\mu$A, $R$ = 1.525 M$\Omega$, low resistance of the
spin valve 100 M$\Omega$ and high resistance 150 M$\Omega$. These choices result in
$V_I R G_{\parallel}^{\tt max} +I_{dc}R$ = 110.19 mV and $V_I R G_{\parallel}^{\tt 1-mismatch} +I_{dc}R$ =
69.67 mV. Thus, even at room temperature, we can detect a single bit mismatch between the two streams with a probability
exceeding 99.9999\% (since 10$^5$ switching trajectories were simulated).

The stochastic LLG equation also allows us to determine the switching time and the energy dissipated during
switching of any magnetostrictive nanomagnet (whether it is the MTJ's soft layer
during decision making, or the spin valve's contacts during writing of reference and input bits)
in the presence of room-temperature
thermal noise \cite{kuntal, Biswas2014, Scholz2001, Lyutyy2015}. Fig. 4 shows the room-temperature
magnetization dynamics, i.e. the magnetization orientation (represented by the angle
$\theta$ that the magnetization vector subtends with a nanomagnet's major axis) as a function of time
after a voltage $V = V_0$ = 110 mV is applied to a pair of electrodes to generate stress in a magnetostrictive nanomagnet and
initiate switching. The voltage is applied abruptly at time $t$ = 0 and withdrawn abruptly at time $t$ = 1.2 ns.
The stress duration of 1.2 ns ensures that all 10$^5$ trajectories simulated switch successfully (switching error probability $<$ 10$^{-5}$).
For each trajectory, the magnetization vector starts out from around the stable state at $\theta = 45^{\circ}$
and switches to a state that is around the stable state at $\theta = 135^{\circ}$.
Because of thermal noise, there is some fluctuation of the initial state around
$\theta = 45^{\circ}$ and the final state around $\theta = 135^{\circ}$.
The temporal characteristics for arbitrarily picked 1000 switching
trajectories
are plotted in Fig. 4. The maximum time it takes for any trajectory to
switch is 1.15 ns.

\begin{figure}[!ht]%
 \centering
  \includegraphics[width=3.5in]{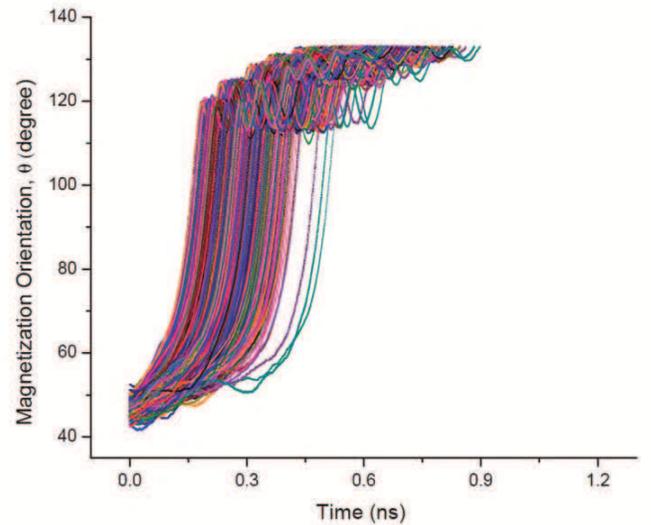}
 \caption[Caption of figure]{Switching dynamics of a nanomagnet in the presence of room-temperature
 thermal noise. A voltage is applied to the electrodes abruptly at time $t$ = 0 and withdrawn abruptly
 at time $t$ = 1.2 ns. The initial orientation of the magnetization is near one stable state
 at $\theta$ = 45$^{\circ}$ and the final orientation is around the other stable state at
 $\theta$ = 135$^{\circ}$. There is some fluctuation around these states owing to thermal noise.
 All 10$^5$ trajectories simulated switch within 1.15 ns. The plot shows arbitrarily picked
 1000 trajectories. Here, thermal fluctuations are not considered after switching.}
\end{figure}

The energy dissipated during switching has two contributions -- internal dissipation due to Gilbert damping in the nanomagnet
and $(1/2)CV^2$ dissipation associated with charging and discharging the electrode capacitances non-adiabatically. The former
contribution is found to be 1.45 aJ (after averaging over 10$^5$ switching trajectories) \cite{Lyutyy2015}
and the latter is 10.64 aJ since $V = V_0$ = 110 mV and $C$ for each of the two electrodes is
 0.88 fF based on electrode area of 100 nm $\times$ 100 nm and PZT layer thickness of
100 nm (the relative dielectric constant of PZT is roughly 1,000). Therefore,
the total energy dissipated to program a reference bit, or write an input bit,
 is $\sim$12.1 aJ. In a 16-bit comparator, the
energy dissipated to program 16 reference bits (or write 16 input bits) will be 16 $\times$ 12.1 aJ = 193.6 aJ.
Even if a reference bit (or input bit) does not change from the previous cycle, we will not know that apriori since we
do not read the stored bit. Therefore, we will have to rewrite every bit and thus all 16 reference bits will have to
be reprogrammed and all 16 input bits will have to be rewritten.

The energy dissipated to switch the soft layer of the MTJ (and hence make its resistance switch) is obviously
also $\sim$12.1 aJ. The time to charge the capacitance in the MTJ unit is $\tau$ = R$_{eq}$C, where
$R_{eq} = R \parallel r_1 \parallel r_2 \parallel \cdots \parallel r_{16} $ = 1.22 M$\Omega$. Hence, the RC 
time constant associated with charging turns out to be 1 ns.

In addition to the energy dissipated during switching, there is dissipation
caused by the current and voltage sources in the spin valve resistances $r_n$, the bias resistance $R$,
and the MTJ resistance $R_{MTJ}$. We can turn on the current and voltage sources
only when
an input bit stream arrives in order to avoid standby energy dissipation. We will keep $V_I$ and $I_{dc}$ on for
the duration $t_{ON}$ = 1.2 ns
and $I_0$ on for the switching duration $t_s$ to ensure both correct decision making and 
correct reading of the decision with $>$ 99.9999\% probability.
The energy dissipated in the $n$-th spin valve is at most $( (V_I-V_0)^2/r_n^{\tt low})t_{ON}$ = (9.89 V)$^2$ $\times$ 1.2 ns /100 M$\Omega$ =
1.17 fJ during the comparison operation and so in a 16 bit comparator, this dissipation is 18.7 fJ.
The energy dissipated in
the resistor $R$ is $(V_0^2/R)t_{ON}$
= (110 mV)$^2$ $\times$ 1.2 ns /1.525 M$\Omega$ = 9.5 aJ. The energy dissipated in the MTJ is $I_0^2 R_{MTJ} t_s$
which can be made arbitrarily small by making $I_0$ arbitrarily small. Therefore in a complete cycle
consisting of programming all 16 bits, receiving 16 inputs and then comparing the programmed bits with the input bits,
 the total energy dissipation is 193.6 aJ + 193.6 aJ + 12.1 aJ + 18.7 fJ + 9.5 aJ = 19.1 fJ. The total time
required to program 16 bits while receiving 16 input bits at the same time (and also ``resetting'' the MTJ at the same time) is 1.2 ns, the time required to produce
the decision (match or no match) by switching the MTJ resistance after the comparison is over is another 1.2 ns
(to ensure that the error probability is less than 10$^{-5}$), and the additional capacitor charging time is 1 ns, resulting in a total delay of
3.4 ns. Therefore, the maximum operating speed is 1/3.4 ns = 294 MHz.

In conclusion, we have proposed and analyzed a spintronic-straintronic reconfigurable N-bit comparator
(which uses spin properties for device functionality and strain to switch the device) and shown that
it is remarkably energy-efficient, relatively error-resilient and reasonably fast at room temperature.
Such comparators have the additional advantage that the result of the comparison can be stored indefinitely in the resistance
state of the MTJ, resulting in non-volatility.

This work was supported by the US National Science Foundation under grants ECCS-1124714 and CCF-1216614. J. A. would also like to acknowledge the NSF CAREER grant CCF-1253370.

\end{document}